\documentstyle[namedreferences,psfig]{kluwer}
\runningtitle{Triple Helmet Streamers}
\begin{opening}
\title{Helmet Streamers with Triple Structures: \\
Weakly Two-Dimensional Stationary States}
\author{Thomas \surname{Wiegelmann}}
\author{Karl \surname{Schindler}}
\institute{Institut f\"ur Theoretische Physik IV, Ruhr-Universit\"at Bochum, 
D-44780 Bochum, Germany}
\author{Thomas \surname{Neukirch}}
\institute{School of Mathematical and Computational Sciences,
University of St. Andrews,
St. Andrews, Scotland}

\end{opening}

\begin{document}
\begin{abstract}

Recent observations of the solar corona with the
LASCO coronagraph on board of the SOHO spacecraft have revealed the
occurrence of triple helmet streamers even during solar minimum, which
occasionally go unstable and give rise to particularly
huge coronal mass ejections.
We present a method to calculate (semi-)analytically
self-consistent
stationary configurations of triple
helmet streamers which can serve as input for stability considerations
and dynamical calculations.
The method is based on an asymptotic expansion procedure 
using
the elongated structure of the streamers.
The method is very flexible and can be used in both Cartesian and spherical
geometry. We discuss the effects of magnetic shear, gravity and
field-aligned flow on open field lines. Example solutions illustrating
the influence of each of these features on the solution structure are
presented.
\end{abstract}

\section{Introduction}
Recent observations of the corona
with the
LASCO coronagraph \cite{schwenn97} on
board of the SOHO spacecraft showed that the corona can be highly
structured even during the solar activity minimum. The observations
revealed a triple structure of the streamer belt which was existent
for several consecutive days.
The observations further showed
that these triple structures occasionally go unstable leading to a
seemingly new and extraordinarily huge kind of coronal mass ejection
(global CMEs). Natural questions arising from these observations are
whether the helmet streamer triple structure is directly
connected with or responsible
for the occurrence of global CMEs and what is the physical mechanism of their formation.
The aim of this paper is to provide a first step towards a better
theoretical understanding of these phenomena.

The structure of helmet streamers and their stability has been studied
both observationally and theoretically
for a long time 
(e.g. \citeauthor{pneuman:kopp71}, \citeyear{pneuman:kopp71};
      \citeauthor{cuperman:etal90}, \citeyear{cuperman:etal90};
      \citeauthor{cuperman:etal92}, \citeyear{cuperman:etal92};
      \citeauthor{koutchmy:livshits92}, \citeyear{koutchmy:livshits92};
      \citeauthor{wang:etal93}, \citeyear{wang:etal93};
      \citeauthor{cuperman:etal95}, \citeyear{cuperman:etal95};
      \citeauthor{wu:etal95}, \citeyear{wu:etal95};
      \citeauthor{hundhausen95}, \citeyear{hundhausen95};
      \citeauthor{bavassano:etal97}, \citeyear{bavassano:etal97}).
A natural association seems to exist
between helmet streamer stability and coronal mass ejections (CMEs) and
sometimes CMEs are accompanied by erupting prominences or filaments.
Coronal streamers are also thought to be 
 the source regions of the slow solar wind and the activity processes discussed
above may well contribute to this component of the solar wind.

Multiple streamer structures have already been observed before SOHO,
however mainly during the maximum phase of the solar activity cycle.
Also, observations of the heliospheric plasma sheet seem to indicate 
a multiple current sheet substructure of the plasma sheet itself 
\cite{crooker:etal93,woo:etal95} which has initiated
several theoretical
investigations of the stability of multiple current sheets
\cite{otto:birk92,yan:etal94,dahlburg:karpen95,birk97,birk:etal97,wang:etal97}.

The aim of this work is to present a simple method by which triple
helmet streamer structures can be calculated self-consistently. We want to keep the method simple
enough so that we can obtain largely analytical results 
without rendering our models too unrealistic.
Since the observed structures seem to be quite elongated
(the radial magnetic field component is considerably larger than
the longitudinal magnetic field component), we can
employ a method which has already been applied very successfully to
the equilibrium structure of the magnetotail of the Earth 
\cite{schindler72,birn:etal75}.
The method is based on an asymptotic
expansion.
Here we generalize this method and
show how it can be applied to find solutions with internal multiple
structures. Obvious generalizations of the original magnetospheric
versions to solar applications
are the use of a more suitable coordinate system (spherical rather than
Cartesian), the inclusion of magnetic shear, of
the solar gravity field for large scale structures and of plasma flow on open field lines.
For reasons of mathematical simplicity and since we only deal with the
subcritical part of the flow, we confine the discussion to  incompressible flow. Our models provide a flexible
quantitative description of structured helmet streamers. They can
also be used as the starting configurations for future studies of
the dynamical evolution of such structures.

The outline of the paper is as follows. In Sect. \ref{sec:method} we present
the mathematical basis of the method. Section \ref{sec:results} contains
representative results obtained by applying the method under various
sets of assumptions. In Sect. \ref{sec:discussion} we summarize the paper,
give a discussion of our results and an outlook on future work.

\section{Mathematical Formulation}
\label{sec:method}

\subsection{Basic equations}

We use the equations of stationary ideal magnetohydrodynamics
to describe the coronal plasma:
\begin{eqnarray}
-\nabla P + \frac{1}{\mu_0} (\nabla \times {\bf B}) \times {\bf B} -
\rho \nabla \Psi &=& \rho {\bf v}\cdot\nabla  {\bf v}     \label{stat} \\
\nabla\cdot (\rho{\bf v}) &=& 0                           \label{cont} \\
{\bf E} + {\bf v} \times {\bf B} & =& 0                   \label{ohm}  \\
\nabla \cdot {\bf B} &=&0                                 \label{divB} \\
P &=& \rho R T  .                                         \label{idealgas}
\end{eqnarray}
Here, $P$ stands for the plasma pressure, 
$\bf B$ for the magnetic field, $\rho$ for the
plasma density,
$\Psi$ for the solar gravitational potential, $\bf v$ for the plasma velocity,
$\bf E$ for the electric field,
$R$ is the gas constant,
$T$ the temperature and $\mu_0$ the vacuum permeability.
Due to the high conductivity of the coronal plasma
the magnetic field is frozen into it (Eq. \ref{ohm}) and we can make the assumption that the observed plasma features outline the magnetic field structure. Throughout the rest of the paper we make the following assumptions:
\begin{itemize}
\item The streamer structure is very extended in azimuth and varies only very
slowly along this direction (this assumption is supported by the observations
\cite{schwenn97} which showed that the triple structure existed for several
days); we can then describe the structure as approximately two-dimensional in the sense of rotational invariance.

\item The streamers are elongated in the radial direction and the radial magnetic field component is considerably stronger than the latitudinal component; this
is the basic assumption allowing to apply an asymptotic expansion procedure.

\item If flow is included, it is purely field-aligned and confined to the
open field line region; since we describe only the low lying parts of the streamer
structure the flow velocities should be subcritical and we will therefore
only consider incompressible flow.

\item For simplicity, we consider only isothermal equilibria; in principle,
other temperature structures or even the inclusion of an energy equation
into the scheme is possible.
\end{itemize}
In the following, we normalize the magnetic field 
by a typical value $B_0$, the plasma pressure $p$  by
$B_0^2/\mu_0$ (where $p=1$ is the maximum value of $p$),
the mass density $\rho$ by $B_0^2/\mu_0 \mbox{R} T$,
the length $L$ by a solar radius and the current density
by $j_y=B_0/\mu_0\,L$. 

\subsection{Method}

To demonstrate the method we first use a two-dimensional
Cartesian geometry with no spatial variation in the $y$-direction ($\frac{\partial}{\partial y}=0$) and assume static structures (${\bf v}=0$)
without magnetic shear ($B_y=0$).
In other geometries the basic properties of the method stay similar though the details may vary due to the geometry.

Writing the magnetic field as
\begin{equation}
{\bf B}(x,z) = \nabla A(x,z) \times {\bf e}_y 
\label{BvonA}
\end{equation}
we find from equation (\ref{stat}) that the flux function $A$ has to obey
the equation
(e.g. \citeauthor{low75} \citeyear{low75};
      \citeauthor{birn:etal78} \citeyear{birn:etal78};
      \citeauthor{low80} \citeyear{low80};
      \citeauthor{birn:schindler81} \citeyear{birn:schindler81})
\begin{equation}
-\Delta A = 
\frac{\partial}{\partial A}\left(
 P(A,\Psi) \right)  \label{gsg1}
\end{equation} 
With the assumption of a constant temperature and because 
$\Psi=\Psi(z)$  in Cartesian geometry we get:
\begin{equation}
P(A,\Psi)=P(A,z)= k(z)\, p(A) \label{grav1}
\end{equation}
with $k(z)=\exp(-(\Psi(z)-\Psi(0))/\mbox{R}T)$. 
For slow variation of $\Psi(z)$ with height $z$ one can neglect gravity
and finds $k(z) = 1$.

If we assume that the magnetic field
perpendicular to the photosphere ($B_z$ in the Cartesian geometry) is much larger than the parallel component ($B_x$ in the Cartesian geometry), 
or in other words, that streamers are rather elongated in the radial direction,
we can use the
method of asymptotic expansion and apply it to Eq. (\ref{gsg1}).
Mathematically stretched configurations are characterized by
the ordering
\begin{equation}
\frac{\partial}{\partial x}=O(1) \Rightarrow B_z=O(1) \qquad 
\frac{\partial}{\partial z}=O(\epsilon) \ll 1  \Rightarrow B_x=O(\epsilon)
\end{equation}
If we neglect terms of the order  $\epsilon^2$ in Eq. (\ref{gsg1})
we obtain after one integration with respect to $x$ :
\begin{equation}
\frac{\partial A}{\partial x}=\pm \sqrt{2 \left(p_0-p(A)\right)} \label{gsg2}
\end{equation}
where $p_0=p_0(z)$ is an integration constant (total pressure on the $z$-axis), which depends parametrically on $z$.
We solve this differential equation by separation \cite{birn:etal75}:
\begin{equation}
x-x_0 = \int_{A_0}^A \frac{d A}{\sqrt{2\left(p_0-p(A)\right)}} \label{gl8}
\end{equation}
and fix the constant $A_0=A(x_0)$ so that 
$p(A_0)=p_0$.

Unfortunately it is impossible to derive the pressure function $p(A)$ directly
from the observations presently available. We will therefore choose  $p(A)$ by a compromise
between physical reasoning and mathematical simplicity. We do not choose
pressure functions which have a too simple dependence on $A$ like
$p(A) \propto \mbox{const.}$, $A$ or $A^2$ because solutions of this type
are a) known to be linearly stable and b) have already been discussed in
\citeauthor{wiegelmann97} \shortcite{wiegelmann97}.
A popular and convenient choice is $p(A) \propto \exp(c A)$ 
because it corresponds to the physical assumption that the plasma is in local thermodynamic equilibrium
and it has the mathematical advantage that one can find analytical solutions.
Using
\begin{equation}
p(A)=k \exp(c A) \label{druck}
\end{equation}
where $k$ may be a function of $z$ and $c$ may be positive or
negative, we get 
from equation (\ref{gl8}):
\begin{equation}
A(x,z) = -\frac{2}{c} \log \left\{
\cosh\left[
           c  \sqrt{\displaystyle\frac{p_0}{2}}(x-x_0)\right]
                                                       \right\}
          +\frac{1}{c}\log\left(\frac{p_0}{k}\right) \label{Axz}
\end{equation}
The components of the magnetic field are then given by
\begin{equation}
B_z=\frac{\partial A}{\partial x}
   = -\sqrt{2 p_0}\tanh \left(c \sqrt{\frac{p_0}{2}} (x-x_0) \right)
\label{Bz}
\end{equation} 
\begin{equation}
B_x= -\frac{\partial A}{\partial z}
  =-\frac{\partial A}{\partial p_0}\frac{\partial p_0}{\partial z}
    -\frac{\partial A}{\partial k}\frac{\partial k}{\partial z}
 \label{Bx}
\end{equation}
We remark that one  cannot evaluate $\frac{\partial A}{\partial p_0}$ and $\frac{\partial A}{\partial k}$ in equation (\ref{Bx}) directly because
in general
$x_0$ depends on  $p_0(z)$ and $k(z)$ as well.
The plasma pressure can be obtained from the Eq. (\ref{druck}) 
\begin{equation}
p=\frac{p_0}{\cosh^2 (\sqrt{\frac{p_0}{2}}c (x-x_0))} \label{pxz}
\end{equation}
and the current density $j_y$ from  $j_y=\frac{\partial p}{\partial A}$:
\begin{equation}
j_y=c p = \frac{p_0 c}{\cosh^2 \left(\sqrt{
\displaystyle\frac{p_0}{2}}c (x-x_0)\right)} .\label{jxz}
\end{equation}
%
\subsection{Triple  structures}

The method just outlined  
has been applied very successfully to calculate the structure
of the Earth's magnetotail
\cite{birn:etal75,schindler:birn82,wiegelmann:schindler95}. 
However, since we want to apply this method to describe
multiple streamer structures a few details of the method have to be changed
though the basic approach stays the same. 

In an arcade structure the direction of the $B_z$ component of the 
magnetic field changes its sign as one crosses the centre of the arcade.
In a triple structure, one will therefore encounter three such changes as 
one moves across the structure. 
In a stretched configuration described by the
asymptotic expansion procedure, the only possibility to achieve this
is  that the current density $j_y$ 
changes its sign between the streamers, or, in other words, in the central streamer the current
flows in the opposite direction with respect to the two outer streamers. 
In terms of the sign of $B_z$ this means that in order to paste the 
three arcades continuously together, the $B_z$ component in the central
streamer has to change just in the opposite sense to the outer two streamers.

Since the current density $j_y$ is directly
linked to the plasma pressure by the relation
$j_y=\partial P/ \partial A$, this implies that the derivative of the
pressure with respect to the flux function $A$ has to change its sign, too.
We can therefore not use a single 
exponential function to model the pressure because the exponential function
and its derivative have no zero.
The simplest alternative is that we use different pressure functions 
for the middle ($p_1(A)$ say) and 
for the outer streamers ($p_2(A)$). In principle, we could also use a different
pressure function for each of the three streamers, but since we do not
want to make the model too complicated we assume
the structure to be mirror symmetric with respect to the $z$-axis
for simplicity. 

Due to the different pressure functions our solution will
have two separatrix field lines 
characterized by $A=A_s$ which separate the regions
of different pressure from each other. These separatrices have to 
be calculated self-consistently together with the solution but due to  our symmetry assumption
we only have to calculate one of those separatrix field lines. 
In the asymptotic expansion procedure the location of the separatrix field line
will be given by $x_{sep}=f(z)$ where $f$ is only varying weakly with
$z$. 

If we require a smooth transition of magnetic field
across the separatrix from the outer to the inner streamer, 
the pressure functions must be continuous. Therefore we choose
\begin{equation}
p_1(A)=k_1 \exp(-c_1 A)
\end{equation}
for the middle streamer and
\begin{equation}
p_2(A)=k_2 \exp(c_2 A)
\end{equation}
for the outer streamers,
while the coefficients $k_1$, $k_2$, $c_1$ and $c_2$ ($c_1, c_2 >0$)
are related by
\begin{equation}
k_2=k_1 \cdot \exp(-(c_1+c_2) A_s) 
\end{equation}
on the separatrix labeled by $A=A_s$.
This choice has the consequence that the current density 
$j_y$ has a discontinuity at the separatrix.
We remark, however, that the absolute value of 
$j_y$ is only rather small at the separatrix and only
changes from a small negative to a small positive value and therefore 
the discontinuity of the current density should not influence
the solution structure very much.

\begin{figure}

\psfig{figure=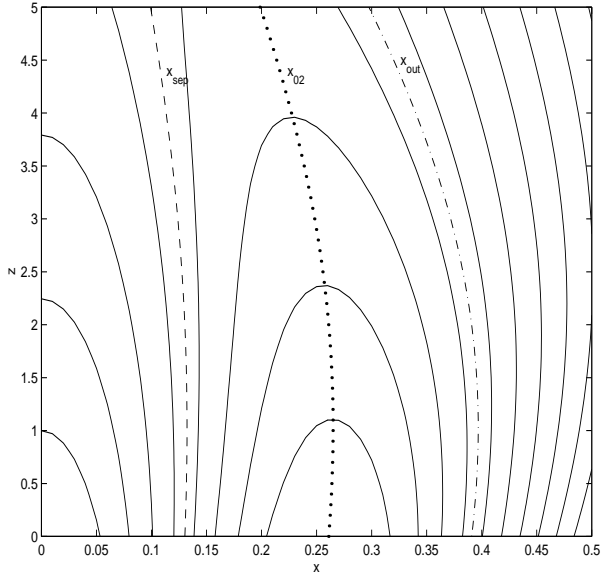,height=8cm,width=8cm}

\caption{This picture illustrates the right side of our configuration
(the left hand side can be derived from the symmetry assumption). The dashed
($--$) line corresponds to the separatrix $x_{sep}(z)$ between the
streamers, the dotted ($\cdots$) line corresponds to the center of the
outer streamer $x_{02}(z)$ and the dashdotted ($-\,\cdot$) line corresponds to the outer separatrix $x_{out}(z)$.}
\label{skizze}
\end{figure}
%
We choose $x_0=0$
to put the centre of the middle streamer on the $z$-axis. 
To calculate the separatrix 
between the middle and one of the outer streamer we use equation 
(\ref{gl8}) with the form of $p(A)$ inside the middle streamer to obtain
 \begin{equation}
x_{sep}=\sqrt{\frac{2}{p_0}} \frac{1}{c_1} 
\mbox{arctanh} \left(\sqrt{\frac{(p_0 - k_1 \exp(-c_1 A_s)}{p_0}} \right) .
\label{sep1}
\end{equation}
Note that both $p_0$ and $k_1$ may depend weakly on $z$.

If we call the centre of the outer streamer $x_{02}$,
we can calculate this central field line relative to the separatrix 
field line again with equation (\ref{gl8}), but
this time using $p_2(A)$ and setting $x_0=x_{02}$:
\begin{equation}
x_{02}=x_{sep}+\sqrt{\frac{2}{p_0}} \frac{1}{c_2} \mbox{arctanh} \left(\sqrt{\frac{(p_0 - k_2 \exp(c_2 A_s)}{p_0}} \right)
\label{sep2}
\end{equation}
If one takes into consideration that $p_1(A_s)=p_2(A_s)$ 
and Equation (\ref{sep1}) one finds:
\begin{eqnarray}
x_{02} &=& \mbox{arctanh} \left(\sqrt{\frac{(p_0 - k_1 \exp(-c_1 A_s)}{p_0}} \right) \sqrt{\frac{2}{p_0}} \left(\frac{1}{c_1} +\frac{1}{c_2} \right) \label{x02} \\ 
\nonumber 
&=& \mbox{arctanh} \left(\sqrt{\frac{(p_0 - k_2 \exp(c_2 A_s)}{p_0}} \right) \sqrt{\frac{2}{p_0}} \left(\frac{1}{c_1} +\frac{1}{c_2} \right) 
\end{eqnarray}
A sketch of the whole structure is shown in Figure \ref{skizze}. 

We can now calculate $A$, $B_z$, $B_x$,
$p$ and $j_y$ as functions of $x$ and $z$
from the Equations (\ref{Axz}),  (\ref{Bz}),  (\ref{Bx}),
(\ref{pxz}) and  (\ref{jxz}) and by using the symmetry properties
of these quantities.
As an input we need the total pressure $p_0(z)$ 
which is the same in all three streamers, the parameters $c_1$, $c_2$ and
$k_1$ and $k_2$. If gravitation is included, $k_1$ and $k_2$ may depend
weakly on $z$, but could be different if we allow different temperatures inside
the streamers.
\subsection{Inclusion of flow}
The method as outlined so far neglects that coronal streamers are the
source regions of the slow solar wind. It is an open question if 
a stationary slow solar wind component exists or whether the
slow solar wind is the result of many small eruptions. Recent
observations (Schwenn and Inhester, private communication 1997)
seem to favour the second alternative. 
If that is true, a steady state model providing the start configuration for a time-dependent computation of the eruption
process need not include the flow, at least not near the symmetry plane.
Nevertheless we want to show
how a stationary plasma flow can be included in our model. 
It may be relevant for the open field regions at larger distances
from the ecliptic, to which we confine the flow.

The inclusion of plasma flow into the asymptotic expansion
procedure has been carried out for both incompressible \cite{geb1,birn91} and
compressible flow \cite{birn92,young:hameiri92}. 
There are three critical points in the solar wind: The slow
magnetosonic point $v_s$, the Alfv\'en point $v_A$ and the fast
magnetosonic point $v_f$ with $v_s < v_A <v_f$ for the
coronal plasma. If one assumes that the flow velocity at the basis
of the corona is less than $v_s$ the solar wind has to pass through all these critical points.   \citeauthor{birn92} \shortcite{birn92} showed that in stretched configurations the total pressure function $p_0(z)$ is related to
the flow velocity. For $v<v_s$ the flow velocity decreases with decreasing
$p_0(z)$ and for $v>v_s$ the flow velocity increases with decreasing $p_0(z)$.
As it is necessary that $p_0(z)$ decreases with respect to $z$ to
get closed arcade structures, the solar wind velocity should have a minimum
at the helmet streamer cusp. This result is of course only valid if
a stationary slow solar wind exists.
The assumption of incompressible flow in the solar wind, however,
cannot be justified in a strict sense, because generally gravity is
not negligible. Since in the paper our main emphasis is placed on
closed structures without directed flow, this model assumption does
not influence our conclusions significantly. (In fact, all our
explicit examples are structures without flow.) We note that the above
mentioned variation of the solar wind velocity remains qualitative
valid, if one neglects gravity and assumes incompressible flow. 
We will assume that the
flow is purely field-aligned and since we will not extend our solution to
the first critical point, we will
for simplicity make
the assumption of the flow being incompressible and 
subalfvenic.

We mention that, as a useful nice property, such stationary solutions with
flow, can be directly calculated by transformation from
static solutions \cite{geb2}.
Here we follow the approach developed in \citeauthor{birn91} \shortcite{birn91}.
Equation (\ref{gsg2}) is then substituted by: 
\begin{equation}
\frac{\partial A}{\partial x}=\pm \sqrt{ \frac{2  \left(p_0-p(A)\right)}{1-M_A^2(A)}} \label{gsg3}
\end{equation}
where $M_A(A)$ is the Alfv\'en Mach number defined by $M_A=v/v_A$, 
where $v_A=B/\sqrt{\mu_0\rho}$
is the Alfv\'en velocity. Note that for incompressible flow both the
Alfv\'en Mach number and the plasma density are constant along open field lines.

To extend the method outlined so far to include flow
on open field lines, 
we have to define a second separatrix field line $A_v$ between the regions
of open and closed magnetic flux. 
Since there shall be no flow in the closed field line regions, we choose
$M_A(A)=0$ on the left hand side of the separatrix and 
$0<M_A(A)<1$ (subalfv\'enic flow)
on the right hand of the separatrix. 
In the special case $M_A=\mbox{constant}$ we find with
equation (\ref{gsg3}) that one can easily include incompressible subalfvenic  plasma flow
($M_A<1$), if we replace the term $(x-x_0)$ in the static theory by $(x-x_0)/\sqrt{1-M_A^2}$. 
The location of the separatrix field line $A_v$ could be calculated by a
similar procedure as the inner separatrix, but
it is more convenient to choose it to be at the same distance from the centre
of the outer streamer as the inner separatrix so that we have a symmetric situation again and one finds $A_v=A_s$. 
We remark that this condition is necessary
if one investigates configurations with a cusp structure.
One can then calculate the shape of the
separatrix defined by $x_{out}(z)$ simply from 
$x_{out}(z)= 2 x_{02}(z)-x_{sep}(z)$ 
(see Figure \ref{skizze} for illustration).

\subsection{Inclusion of magnetic shear}
It is possible to extend our method and include magnetic shear writing ${\bf B}$ as:
\begin{equation}
{\bf B}(x,z) = \nabla A(x,z) \times \vec e_y + B_y(x,z) \vec e_y \label{BvonA2}
\end{equation}
Using the $y$ component of equation (\ref{stat}) one finds $B_y=B_y(A(x,z))$ and instead of (\ref{gsg1}) we get:
\begin{equation}
-\Delta A = 
\frac{\partial}{\partial A}\left(
\Pi(A,\Psi) \right)  \label{scher1}
\end{equation} 
with $$\Pi(A,\Psi)=\Pi(A,z)=k(z) p(A) +  \frac{B_y^2(A)}{2}.$$
Using the method of asymptotic expansion we get:
\begin{equation}
x-x_0 = \int_{A_0}^A \frac{d A}{\sqrt{2\left(p_0(z)-\Pi(A,z)\right)}} \label{scher2}
\end{equation}
We remark that the reason for magnetic shear should be a 
displacement of magnetic foot points and that only
closed field lines will be sheared. 
We assume $B_y(A) \propto \exp(d A)$ for mathematical simplicity. In general we cannot solve
equation (\ref{scher2}) analytically. This is only possible if
we choose the special case $d=\frac{c}{2}$ which we will do to
illustrate the effect of magnetic shear. 
Then we get:
\begin{equation}
 \Pi(A,z)=\left(k(z) + \frac{\lambda^2}{2}\right) \exp(c A)
\end{equation}
and we can easily include magnetic shear by substituting
$k(z)$ by $k(z)+\frac{\lambda^2}{2}$ in equation (\ref{Axz}).
It is possible to investigate other values ($c \not= 2 d$)
numerically, but one can see the main effect of magnetic shear already in this special case.

We calculate triple structures analogous to the method without
shear and substitute $p(A,z)$ by $\Pi(A,z)$.  We have to consider shear only on closed field lines. We use
$\Pi_1(A)=\left(k(z) + \frac{\lambda^2}{2}\right) \exp(-c_1 A)$ in
the middle streamer, $\Pi_2(A)=\exp(-(c_1+c_2)A_s) \left(k(z) + \frac{\lambda^2}{2}\right) \exp(c_2 A)$ in the outer streamers and
$\Pi_3(A)=\exp(-(c_1+c_2)A_s) k(z) \exp(c_2 A)$ on open field lines.
Thus there is a jump in $\Pi$ and a corresponding jump in the magnetic field at the boundary between outer streamers and open field lines and consequently there is  a thin current sheet.  Unlike the calculations on
closed field lines one cannot fix $\Pi_3(A_{0})=p_0$. Instead of
(\ref{gl8}) one has to use:
\begin{equation}
x-x_0=\int_{A_{v}}^{A} \frac{d A}{\sqrt{2(p_0-\Pi_3(A))}}
\end{equation}
where $A_{v}$ is the value of the last closed field line
(boundary field line between open and closed regions). One finds
$A_v=A_s$ for configurations with cusp structure.

We remark that violating the condition $\Pi_1(A_s)=\Pi_2(A_s)$
could still make sense. For example, one could use 
$P_1(A_s)=P_2(A_s)$ but 
$B_{y1}(A_s) \not= B_{y2}(A_s)$. 
Of course this mismatch in the value of $\Pi$ would have to be
compensated by a corresponding mismatch in $B_z$ and this would lead to additional
current sheets between the middle and the outer streamers.


\subsection{Spherical coordinates}
In this subsection we formulate the method in spherical coordinates
 $(r,\theta,\phi)$, which are more realistic to describe the solar corona
on scales comparable to or larger than a solar radius. 
We assume static $({\bf v}=0)$ configurations and do not
consider spatial variation in the $\phi$-direction $(\frac{\partial}{\partial \phi}=0)$. Thus we can present the magnetic field as:
\begin{eqnarray}
B_r & =& \frac{1}{r^2 \sin \theta} \frac{\partial \xi}{\partial \theta} \nonumber \\
B_{\theta} &=& - \frac{1}{r \sin \theta} \frac{\partial \xi}{\partial r} \nonumber \\
B_{\phi} &=& B_{\phi}(r,\theta)
\end{eqnarray}
Here $\xi$ is the flux function in spherical coordinates.
We remark that in spherical coordinates the flux function
is not identical with a component of the magnetic vector potential. One finds $\xi=r \sin(\theta) A_{\phi}$, where
$A_{\phi}$ is the $\phi$-component of the vector potential ${\bf A}$. 
We find from
(\ref{stat}) that $\xi(r,\theta)$ has to obey the equation:
\begin{equation}
\frac{\partial^2 \xi}{\partial r^2}+\frac{\sin \theta}{r^2}
\frac{\partial}{\partial \theta}\left(\frac{1}{\sin \theta} \frac{\partial \xi}{\partial \theta} \right) = - r^2 \sin^2 \theta \frac{\partial P(\xi,\Psi)}{\partial \xi}-\frac{1}{2} \frac{\partial f^2(\xi)}{\partial \xi} \label{gradkugel}
\end{equation}
where $f(\xi)= r \sin \theta B_{\phi} $ is constant on field lines.
With the assumption of a constant temperature and because 
$\Psi=\Psi(r)$ one finds:
\begin{equation}
P(\xi,\Psi)=k(r) p(\xi)
\end{equation}
with $k(r)=\exp(-(\Psi(r)-\Psi(0))/\mbox{R}T)$. 

We assume that the magnetic field perpendicular to the
photosphere $B_r$ is much larger than the parallel component
$B_{\theta}$. For simplicity we do not consider magnetic shear
here $(B_{\phi}=0)$. Stretched configurations are characterized by
$$
\frac{1}{r^2}\frac{\partial}{\partial \theta}=O(1) \Rightarrow B_r=O(1) \quad \frac{1}{r}\frac{\partial}{\partial r}=O(\epsilon) \ll 1 \Rightarrow B_{\theta}=O(\epsilon) 
$$
This approximation is valid for configurations close to the
equatorial plane ($\theta=\frac{\pi}{2}$) and $r^2 p_0(r)$ has to
vary only slowly with respect to $r$. The second condition
limits the outer boundary of our model corona to a few solar radii.
As the observed configurations
are indeed concentrated near the equatorial plane and the closed streamer field lines seem to be radially bounded to a few (about $2-4$) solar radii, our asymptotic model should describe these configurations approximately correct. 

If we neglect terms of the order $\epsilon^2$ in Eq. (\ref{gradkugel}) we obtain after one integration with
respect to $\theta$:
\begin{equation}
\frac{\partial \xi}{\partial \theta}=^{+}_{-} r^2 \sin(\theta) 
\sqrt{2(p_0(r)-k(r)  p(\xi))}
\end{equation}
We use (analogous to our calculations in Cartesian geometry) $p(\xi)=\exp(c \xi)$ and solve this differential equation by
separation. Instead of (\ref{Axz}) we get:
\begin{equation}
\xi(r,\theta)=\frac{\log(p_0(r))-\log(k(r)-2 \log(\cosh(c \sqrt{\frac{p_0}{2}} r^2 \left(\cos(\theta)-\cos(\theta_0)\right))}{c}
\end{equation}

We remark that one cannot find a one dimensional solution of
equation (\ref{gradkugel}) with $\xi=\xi(\theta)$ like in Cartesian geometry. Thus no analogy to the Harris sheet in Cartesian geometry
exists in spherical coordinates.

To calculate triple structures in spherical geometry one uses
a similar procedure as described for Cartesian geometry.
We choose $\theta_0=\frac{\pi}{2} \Rightarrow \sin \theta_0=0$
to put the centre of the middle streamer and calculate the
separatrix field line analogous to (\ref{sep1}) and get:
\begin{equation}
\cos(\theta_{sep})=r^2 \sqrt{\frac{2}{p_0}} \frac{1}{c_1} 
\mbox{arctanh} \left(\sqrt{\frac{(p_0 - k_1 \exp(-c_1 A_s)}{p_0}} \right) 
\label{ksep1}
\end{equation}
If we call the center of the outer streamer $\theta_{02}$ we get
analogous to our calculations in Cartesian geometry:
\begin{eqnarray}
\cos(\theta_{02}) &=&r^2 \sqrt{\frac{2}{p_0}} \left(\frac{1}{c_1} + \frac{1}{c_2} 
\right)
\mbox{arctanh} \left(\sqrt{\frac{(p_0 - k_1 \exp(-c_1 A_s)}{p_0}} \right) \nonumber \\
 &=&r^2 \sqrt{\frac{2}{p_0}} \left(\frac{1}{c_1}+ \frac{1}{c_2} 
\right)
\mbox{arctanh} \left(\sqrt{\frac{(p_0 - k_2 \exp(-c_2 A_s)}{p_0}} \right)
\label{ksep2}
\end{eqnarray}
The quantities $\xi$, $B_r$, $B_{\theta}$, $p$, $j_{\phi}$ can then be
calculated as
functions of $r$ and $\theta$ in the usual way.
\section{Results}
\label{sec:results}

\subsection{Static solutions in Cartesian geometry without gravitation}

\begin{table}
\caption{Parameter sets used to calculate the solutions presented in 
Figure \protect\ref{fig1}}
\begin{tabular}{lllllll}\hline
Solution & $s_1$ & $s_2$ & $s_3$ & $c_1$ & $c_2$ & $A_s$  \\ \hline
a)       &  0.8  &  0.4 &  0.2  &  15.0  &  15.0  &  0.1   \\
b)       &  0.8  &  0.4 &  0.2  &  15.0  &  15.0  &  0.08   \\
c)       &  0.8  &  0.4 &  0.2  &  20.0  &  20.0  &  0.12   \\
d)       &  0.8  &  0.4 &  0.2  &  10.0  &  30.0  &  0.12   \\
e)       &  0.8  &  0.4 &  0.2  &  30.0  &  12.0  &  0.04   \\
f)       &  0.8  &  1.2 &  0.2  &  15.0  &  15.0  &  0.12   \\ \hline
\end{tabular}
\label{table1}
\end{table}

We first give a few examples for triple streamers 
calculated with our model to illustrate the influence of the
different functions and parameters on the structure of the solutions. 
To calculate a solution we have to specify the functions 
$p_0(z)$ and $k_1(z)$  and the parameters $c_1$, $c_2$ and $A_s$. 
In the first examples we 
neglect the effect of plasma flow and of the solar gravitation which 
gives $k_1(z)=1$. Our solutions are confined within the boundaries
$x=-0.5 .. 0.5$ and $z=0 .. 5$ corresponding to one solar radius in the
$x$ direction and five solar radii in the $z$ direction. We use symmetry with respect to the $z$ axis.

We prescribe the total pressure by $p_0(z)=s_1 \exp(-s_2 z)+s_3$.
Within our model $A(x,z)$ depends only parametrically on
$z$ by $p_0(z)$ and so the exact form of $p_0(z)$ has not too much 
influence on the configuration as long as its general properties are similar. 
One important property of $p_0(z)$ is that it has to be a monotonically
decreasing function of $z$ to obtain singly connected
closed field line regions.
Physically
one can identify the total pressure on the $x$-axis with the sum of
the magnetic pressure $B^2 /2 $ on open field lines outside the helmet streamer and a homogeneous plasma pressure of the solar wind.

\begin{figure}
\leavevmode\centering
\mbox{\psfig{figure=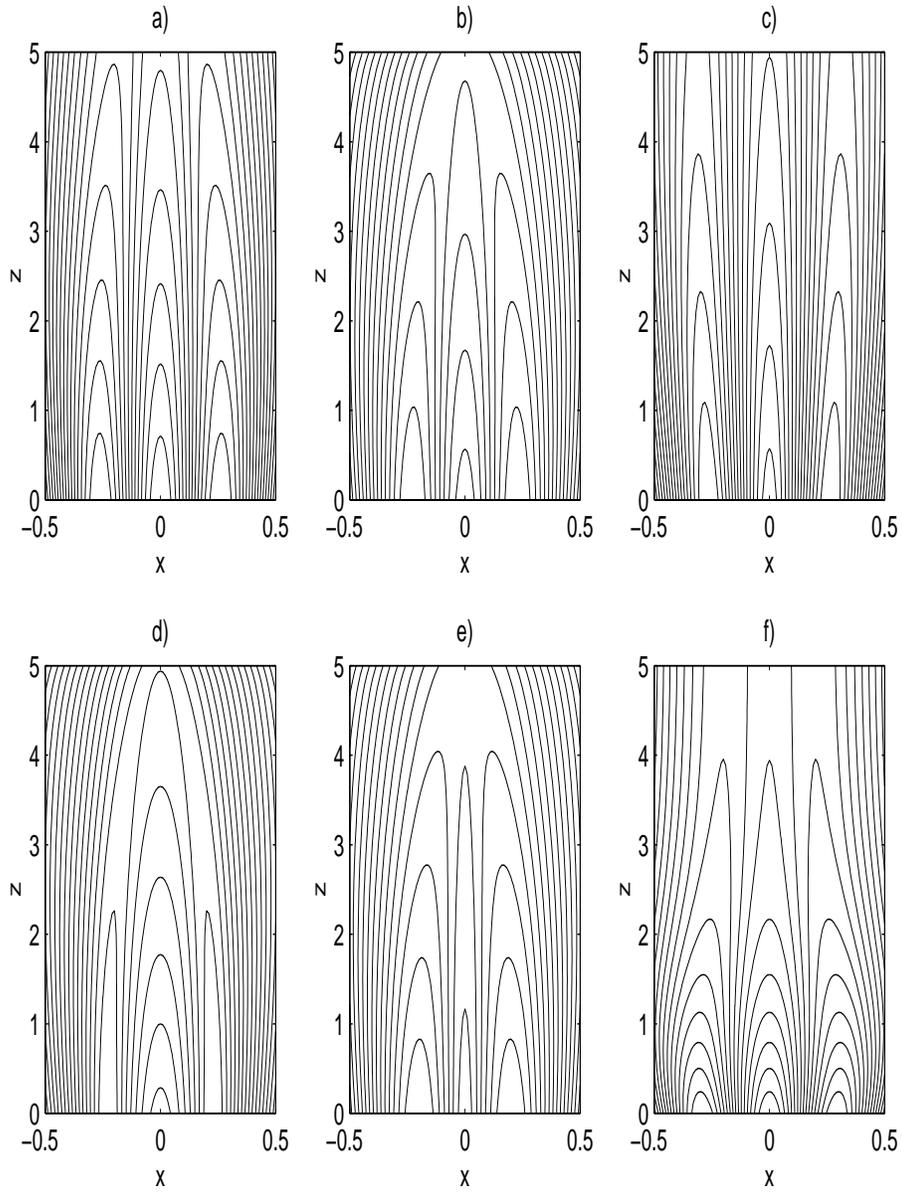,height=16cm,width=12cm}}
\caption{Magnetic field lines (contourplot of $A(x,z)$) for triple helmet streamers (see text)} 
\label{fig1}
\end{figure}
In Figure \ref{fig1} we show field line plots of six solutions calculated
for different sets of parameters which are summarized in Table \ref{table1}.
In Figure \ref{fig1}a) - \ref{fig1}c) we investigate the influence
of the location of the separatrix field line on the solutions structure,
whereas $p_0$ is kept fixed and the parameters $c_1$ and $c_2$ are
fixed to a value of $15.0$ in a) and b) whereas we used $c_1=c_2=20.0$ in c) which leads to thinner configurations. 
In these cases all three streamers have exactly the same width though the
width itself varies. In Figure \ref{fig1}a) the separatrix is a straight
line giving rise to parallel streamers. In Figure \ref{fig1}b) the
separatrix has been moved closer to the centre of the structure and the
three streamers converge, whereas in Figure \ref{fig1}c) the separatrix
has been moved further out giving rise to diverging streamers.

In Figure \ref{fig1}d) we have decreased the value of $c_1$ to $10.0$ and
increased the value of $c_2$ to $30.0$. Also the value for $A_s$ 
has been increased. The effect is that the central streamer now contains more
magnetic flux than the outer streamers and is wider in the $x$-direction. In
Figure \ref{fig1}e) this effect has been reversed by making $c_2$ smaller than
$c_1$. It is obvious that now the outer streamers are wider than the central streamer.

Finally, in Figure \ref{fig1}f) we have varied  the total pressure
$p_0(z)$ by increasing the parameter $s_2$ to $1.2$ resulting in a faster
decrease of $p_0$ with $z$. One notices that the closed field line regions (which we
define here loosely as those field lines not crossing the upper boundary) has
shrunk and that the whole structure has a much slimmer appearance for large
$z$.
\begin{table}
\caption{Parameter sets used to calculate the solutions presented in 
Figure \protect\ref{figcusp}}
\begin{tabular}{llllllll}\hline
Solution & $a$ & $b$ & $n$ & $z_{\mbox{\tiny cusp}}$ & $c_1$ & $c_2$ & $A_s$  \\ \hline
a)       &  0.8  &  0.2 &  3.0  & 4.0 & 15.0  &  15.0  &  0.1073   \\
b)       &  0.8  &  0.2 &  4.0  & 4.0 & 10.0  &  25.0  &  0.1609   \\
c)       &  0.8  &  0.2 &  6.0  & 4.0 & 25.0  &  12.0  &  0.0644   \\
\hline
\end{tabular}
\label{table2}
\end{table}

In a next step we use our model to describe helmet streamer configurations with a cusp. Up to now we have not given a radial
boundary of the  closed magnetic field lines. Now we define such a
point at $x=0$, 
$z=z_{\mbox{\tiny cusp}}$ which defines the last closed field line.
 
\begin{figure}
\leavevmode\centering
\mbox{\psfig{figure=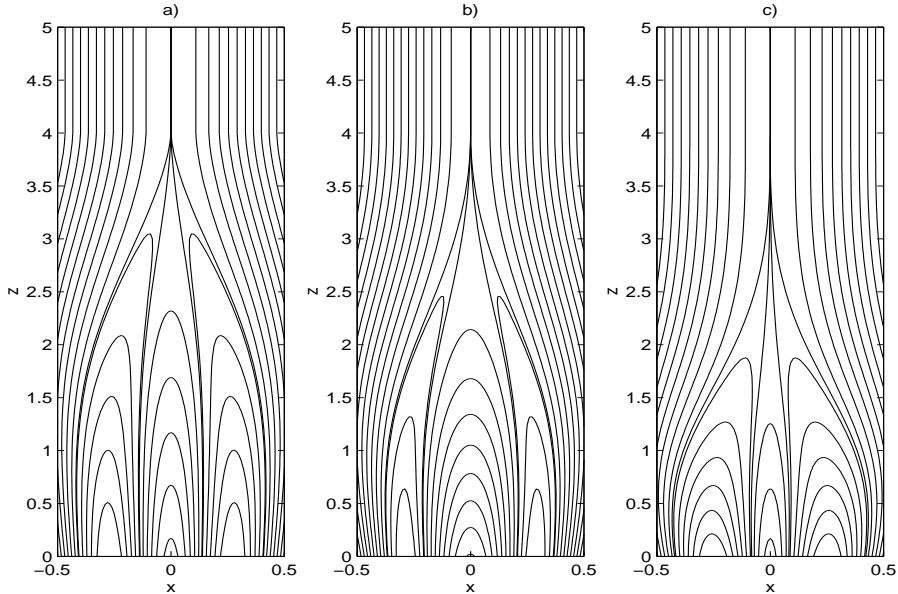,height=8cm,width=12cm}}
\caption{Magnetic field lines (contourplot of $A(x,z)$) for triple helmet streamers with cusp (see text for discussion).} 
\label{figcusp}
\end{figure}

We use as a function for the total pressure: 
\begin{equation}
p_0(z) = \left\{ \begin{array}{r@{\quad\mbox{for}\quad}l}
           a (\frac{z_{\mbox{\tiny cusp}}-z}{z_{\mbox{{\tiny cusp}}}})^n+b  & 
                                            z \leq z_{\mbox{\tiny cusp}}   \\
           b                          &      z>z_{\mbox{\tiny cusp}}
\end{array} \right.
\end{equation}
Field line plots for three different sets of parameters are shown in 
Figure \ref{figcusp}. The parameter sets used are listed in Table \ref{table2}.
It can be seen that the effect of changing the parameters $c_1$ and $c_2$ stays
the same as in the case without cusp. An increase of the exponent $n$ leads
to a stronger decrease of $p_0(z)$ as the cusp point is approached from
below and this leads to a sharper cusp structure.

These examples illustrate the flexibility of our model. One can describe different types of triple helmet streamers by specifying
the free parameters in the general solutions.
\subsection{Static solutions in Cartesian Geometry Including Magnetic Shear and Gravitation}
\label{shear}
\begin{figure}

\leavevmode\centering

\mbox{\psfig{figure=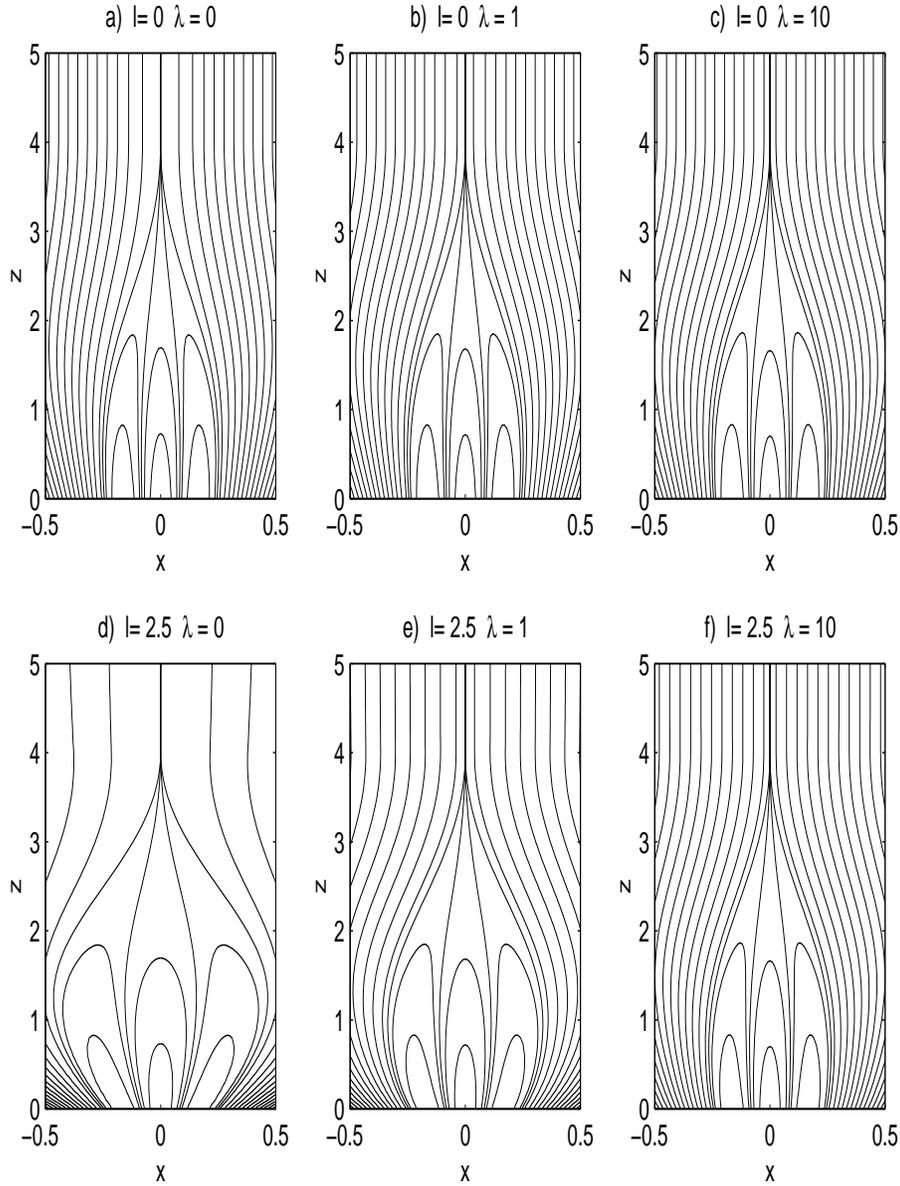,height=16cm,width=12cm}}
\caption{Magnetic field lines (contourplot of $A(x,z)$) for triple helmet streamers. In the upper pictures b) and c) we included magnetic shear and in the lower pictures we additionally
investigated the effect of gravitation.} 
\label{figscher}
\end{figure}
\begin{figure}

\leavevmode\centering

\mbox{\psfig{figure=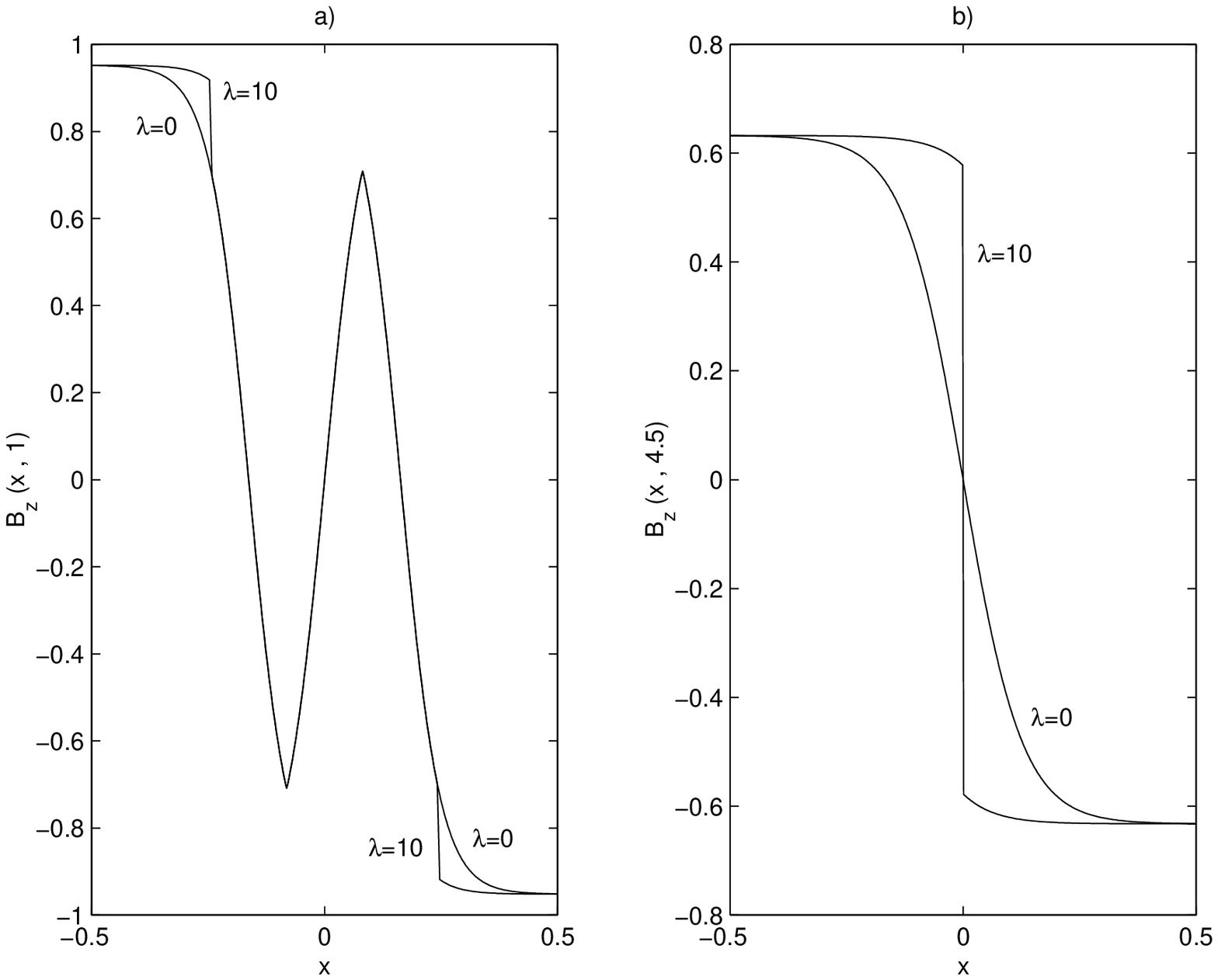,height=6cm,width=12cm}}
\caption{The main magnetic field component $B_z(x)$ without ($\lambda=0$) and with $(\lambda^2=10)$ shear: a) Below the cusp $(z=1.0)$; b) above the cusp ($z=4.5$).} 
\label{figBz}
\end{figure}
In this section we investigate the influence of the solar gravitational field
on the structure of the solutions. Though we are still calculating in
Cartesian geometry we will not take the gravitational force to be constant
as it is usually done in Cartesian geometry if the length scale under
consideration is much smaller than a solar radius (e.g. 
\citeauthor{zwingmann87} \citeyear{zwingmann87}; \citeauthor{platt:neukirch94}
\citeyear{platt:neukirch94}), but use the more appropriate potential
$\Psi=\Psi(z)=-\frac{G M_s}{z+R_s}$.
Together with the assumption of constant temperature we get:
\begin{equation}
k(z)= \exp \left(-\frac{l}{R_s} \right) \cdot \exp \left(\frac{l}{z+R_s} \right)
\end{equation}
where $G$ is the constant of gravitation,
$M_s$ the solar mass, $R_s$ the solar radius
and $l=\frac{G M_s}{R T}$.
Under the assumption of a pure hydrogen plasma and a constant temperature 
 of $3 \cdot 10^6$ K one finds
$l\approx 7.5$ corresponding to $7.5\,R_s$.
One difficulty of the method in Cartesian geometry is that
$k(z)$ must decrease more slowly with respect to $z$ than $p_0(z)$
because otherwise one cannot find
arcade type solutions. This is difficult to achieve with a realistic value
of $l$, but since
we only want to investigate the qualitative
effect of gravitation on the solution we use the unrealistic but 
mathematically more convenient value
$l=2.5$. We remark that this problem is alleviated in spherical geometry which is more realistic anyway (see section \ref{seckugel})

The inclusion of shear only makes sense on closed field lines. As we have
a well defined boundary between open and closed regions only in solutions
with cusp structure, we only investigate such configurations here.
In Figure \ref{figscher} we fix the parameters $c_1=c_2=25.0, z_{cusp}=4.0, a=0.8, n=4.0, b=0.2$ and use for the total pressure:
\begin{equation}
p_0(z) = \left\{\begin{array}{r@{\quad\mbox{for}\quad}l}
            \left(k(z)+\frac{\lambda^2}{2} \right) 
            \left[a
            \left(\frac{z_{\mbox{\tiny cusp}}-z}
                       {z_{\mbox{{\tiny cusp}}}}\right)^n+b\right]  & 
                                            z \leq z_{\mbox{\tiny cusp}}   \\
     b\left(k(z)+\frac{\lambda^2}{2} \right) &      z>z_{\mbox{\tiny cusp}}
\end{array} \right.
\end{equation}
The factor $\left(k(z)+\frac{\lambda^2}{2} \right) $ can be used to fix
the normal component of the magnetic field at the solar surface, which
we could use as a boundary condition. 

The main effect of magnetic shear is that there is a jump in the magnetic field and consequently a thin current sheet between the closed and open field lines. 
Indications of such a current sheet are also found in 
recent observations \cite{schwenn97}. In Figure \ref{figBz}a) we plot
$B_z$  as a function of $x$ at the height $z=1$ for the same parameters
as in Figures \ref{figscher}a) and c).

If we do not include shear the structure above the helmet streamer cusp
$z > z_{cusp}$ becomes equivalent to a Harris sheet and $B_z$ goes smoothly
through zero in the center of the configuration. 
The inclusion of shear on closed field lines leads to a jump
in $B_z$ from a positive to a negative value in the center of the whole configuration (see figure  \ref{figBz} b)).
This
causes a thin current sheet, which is more pronounced than the
current sheet at the outer boundary of the streamer configuration below
the cusp. This current sheet corresponds to the 
heliospheric current sheet. The effect can clearly be seen in Figure \ref{figscher}. In the Figures \ref{figscher}b) and c) we included
magnetic shear and in c) the shear is stronger than in b) leading to
a larger jump in $B_z$. This can be seen in 
the contourplot of $A$ as the field line density is much higher
in the center above the cusp in Figure \ref{figscher}c) than in b). 
In the lower panel (Figure \ref{figscher}d)-f)) we investigated the
same configurations as in the upper pictures but included
gravitation. One can see that without shear (Figure \ref{figscher}d)), 
the configuration gets wider with increasing $z$. (If one uses the
realistic value $l=7.5$ instead of $l=2.5$ the configuration will
become unrealistically wide without shear.) The inclusion of 
shear reduces this
effect and if one investigates force free configurations $\left( \frac{\partial P(A)}{\partial A}=\nabla P=0 \right)$ or nearly force free solutions $\left( \frac{\partial B_y^2(A)}{\partial A} \gg \frac{\partial P(A)}{\partial A} \right) $ which is accomplished in c) and f) the effect of
gravitation vanishes.
\subsection{Solutions in spherical geometry}
\label{seckugel}
Now we present some example solutions of helmet streamer configurations in spherical coordinates. This geometry is of course
more realistic to describe the coronal magnetic field on large scales. 
In this section we also include the influence of the solar gravitational field on the structure of the solutions from the beginning.
With
$\Psi=\Psi(r)=-\frac{G M_s}{r}$ 
we get as in Cartesian geometry (see Section \ref{shear})
\begin{equation}
k(r)= \exp \left(-\frac{l}{R_s} \right) \cdot \exp \left(\frac{l}{r} \right)
\end{equation}
Here we use the realistic value $l=7.5$. 

\begin{figure}
\leavevmode\centering
\mbox{\psfig{figure=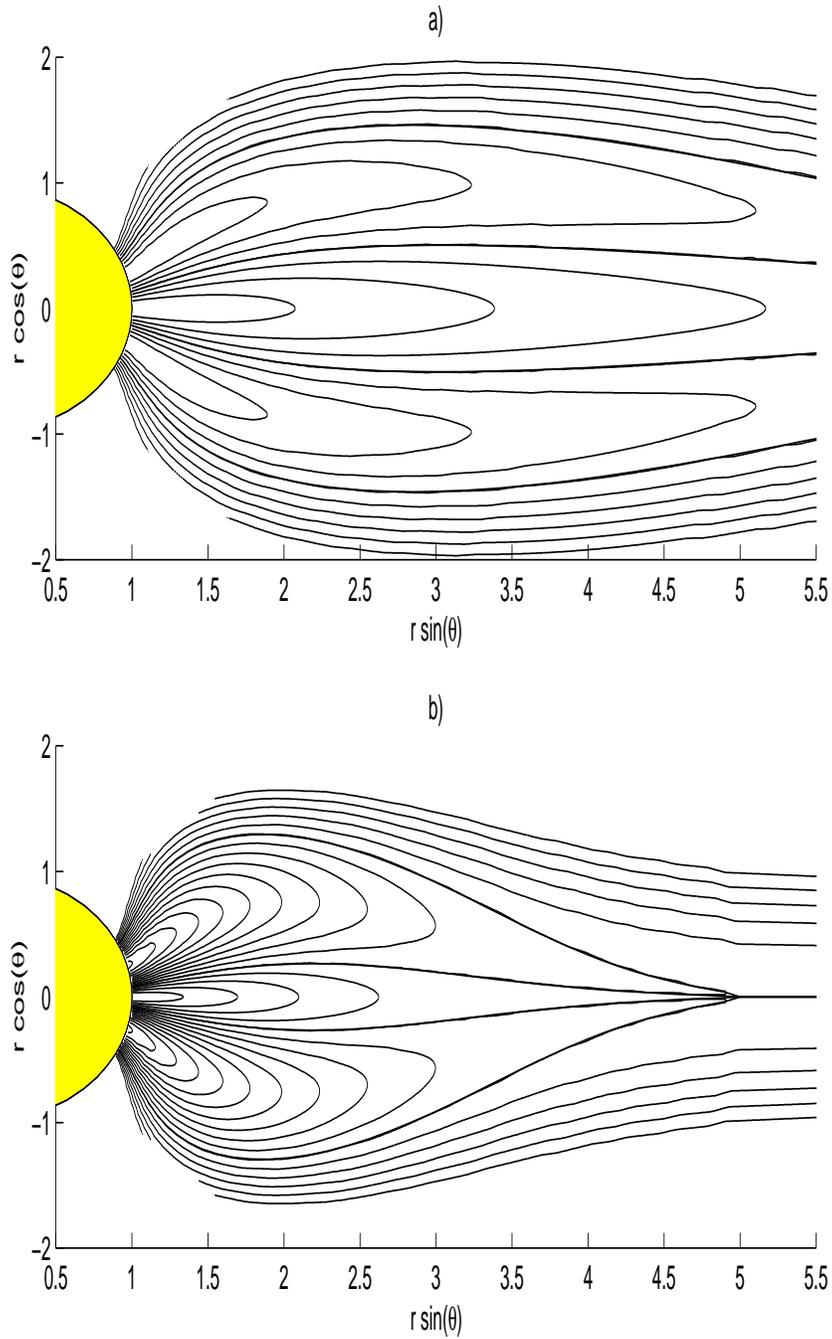,height=18cm,width=11cm}}
\caption{Magnetic field lines (contourplot of $\xi(r,\theta)$) for triple helmet streamers in spherical geometry (see text)} 
\label{figkugel}
\end{figure}
In Figure \ref{figkugel}a) we show a field line plot of a solution in which we
prescribed the total
pressure as $p_0(r)=k(r) \left[s_1 \exp(-s_2(r-1))+s_3 \right]$
and used the same parameters as in Figure \ref{fig1}a) (see Table \ref{table1}, case a). Thus
$p_0(r)/k(r)$ varies in the same way as $p_0(z)$ in the Cartesian
case without gravitation. The effects of the different geometry show up as
a bending of the out streamers towards the equatorial plane.

In Figure \ref{figkugel}b) (see Table \ref{table2}, case b) 
we present a solution with cusp structure in spherical geometry 
with the total
pressure 
\begin{equation}
p_0(r) = \left\{\begin{array}{r@{\quad\mbox{for}\quad}l}
k(r)\left[ a (\frac{r_{\mbox{\tiny cusp}}-r+1}
                   {r_{\mbox{{\tiny cusp}}}})^n+b \right]
                             &   r \leq r_{\mbox{\tiny cusp}}  \\
k(r) b &   r>r_{\mbox{\tiny cusp}} \end{array} \right.
\end{equation}
and use the same parameters as in Figure \ref{figcusp}c). Again
the bending of the outer streamers is obvious.

We remark that prescribing the parameters $c_1, c_2$ and
$p_0(R_s)$ is equivalent to prescribing the normal component of
the magnetic field at the solar surface. This property of the
solutions could be used to match measured magnetic fields and generate
the corresponding streamer structure.

\section{Conclusions and outlook}
\label{sec:discussion}

We have developed a comparatively simple procedure based on the method
of asymptotic expansion
to calculate equilibria
of triple helmet streamer structures. 
The procedure represents a generalization of the 
asymptotic expansion method used in the theory of magnetotail equilibria
\cite{schindler72,birn:etal75,schindler:birn82}. We have 
applied this procedure to the calculation of stationary states of
triple helmet streamers under various conditions and investigated the
dependence of the solution properties on the model parameters, the influence
of the solar gravitational field, the inclusion of field aligned flow and
on the basic geometry used.

As a next step we plan to use the calculated configurations as start
equilibria for MHD simulations. 
These simulations will answer the question of stability which we did
not address in this paper and provide important insights into possible
mechanisms of eruptive phenomena in the
solar atmosphere, especially those connected with a triple structure of the solar corona. The simulations should have relevance both for
huge coronal mass ejections and very small but frequent eruptions which may represent the source mechanism of the slow solar wind. 
\acknowledgements
The authors thank 
Alan Hood, Bernd Inhester, Eric Priest and Rai\-ner Schwenn for discussions
and useful comments.
We also thank the referee, R.A. Kopp, for his useful remarks.
 We acknowledge financial support by the
DFG Graduiertenkolleg ``Hoch\-tem\-pe\-ra\-tur\-plasmaphysik'' (TW), by PPARC (TN)
and by a British-Ger\-man Academic Research Collaboration grant.
\appendix
Here we calculate the terms $\frac{\partial A}{\partial p_0}$ and $\frac{\partial A}{\partial k}$ which where  not
explicitly given in equation (\ref{Bx}).

\subsection*{Middle Streamer}
With $c=-c_1$, $k=k_1$ and $x_0=0$ one finds:
\begin{displaymath}
\frac{\partial A}{\partial p_0}=
\frac{1}{c_1} \left(-\frac{1}{p_0}+\frac{c_1 x}{\sqrt{2 p_0}} \tanh(\sqrt{\frac{p_0}{2}} c_1 x) \right)
\end{displaymath}
\begin{displaymath}
\frac{\partial A}{\partial k_1 }=
\frac{1}{k_1 c_1}
\end{displaymath}

\subsection*{Outer Streamers}
With $c=c_2$, $k=k_2$ and $x_0=x_{02}$ (see equation(\ref{x02}))
one finds:
\begin{eqnarray*}
\frac{\partial A}{\partial p_0} &=&
\frac{\left(2 \sqrt{ p_0} \sqrt{\alpha_1} c_1 
\sqrt{\frac{1}{p_0}} \cosh(\alpha_2 ) 
+ \sinh(\alpha_2) p_0
\sqrt{2} \sqrt{\alpha_1} \sqrt{\frac{1}{p_0}} x c_1 c_2
 - 2 \sinh(\alpha_2) (c_1 + c_2) \right)} 
{2 \left(p_0^{3/2} \sqrt{\alpha_1} c_1 \sqrt{\frac{1}{p_0}}
\cosh(\alpha_2) c_2   \right)}  
\end{eqnarray*}
\begin{eqnarray*}
\frac{\partial A}{\partial k_2}=
\lefteqn{ \left( {\vrule height0.84em width0em depth0.84em}
 \right. \! \! {\rm sinh}(\,{\rm \alpha_2}\,)\,{\it p_0 }\,
\sqrt {{\displaystyle \frac {1}{{\it p_0}}}}\,{\it c_2
} + {\rm sinh}(\,{\rm \alpha_2}\,)\,{\it p_0}\,
\sqrt {{\displaystyle \frac {1}{{\it p_0}}}}\,{\it c_1
}} \\
 & & \mbox{} - \sqrt {{\it p_0}}\,\sqrt {{\rm \alpha_1}}\,
{\it c_1}\,{\rm cosh}(\,{\rm \alpha_2}\,) \! \! \left. 
{\vrule height0.84em width0em depth0.84em} \right)  \left/ 
{\vrule height0.41em width0em depth0.41em} \right. \! \!  \left( 
{\vrule height0.41em width0em depth0.41em} \right. \! \! \,
\sqrt {{\it p_0}}\,\sqrt {{\rm \alpha_1}}\,{\it k_2}\,
{\it c_1}\,{\rm cosh}(\,{\rm \alpha_2}\,) 
{\it c_2}\, \! \! \left. {\vrule 
height0.41em width0em depth0.41em} \right)  \\
\end{eqnarray*}

Therefore we used as abbreviations:
\begin{eqnarray*}
 & &  \alpha_1 := {\displaystyle 
                         \frac{p_0 -  k_2\,\exp(c_2\, A_s)}{p_0}} \\
 & & {\rm \alpha_2} := {\displaystyle \frac {1}{2}}\sqrt {{\it p_0
}} \left( {\vrule height0.84em width0em depth0.84em}
 \right. \! \!  - \sqrt {2}\,{x}\,{\it c_1}\,{\it c_2
} + 2\,{\rm arctanh} \left( \! \,\sqrt {{\rm \alpha_1}}\, \! 
 \right) \,\sqrt {{\displaystyle \frac {1}{{\it p_0}}}}\,
{\it c_2} \\
 & & \mbox{} + 2\,{\rm arctanh} \left( \! \,\sqrt {{\rm \alpha_1}}\,
 \!  \right) \,\sqrt {{\displaystyle \frac {1}{{\it p_0}
}}}\,{\it c_1} \! \! \left. {\vrule 
height0.84em width0em depth0.84em} \right)  \left/ {\vrule 
height0.37em width0em depth0.37em} \right. \! \! {\it c_1}
\end{eqnarray*}

%

\end{document}